\documentclass[10pt,a4paper]{article}

\usepackage[T1]{fontenc}
\usepackage[utf8]{inputenc}
\usepackage{lmodern}
\usepackage[margin=25mm]{geometry}
\usepackage{setspace}
\usepackage[switch]{lineno}
\usepackage{graphicx}
\usepackage{booktabs}
\usepackage{tabularx}
\usepackage{array}
\usepackage{amsmath,amssymb}
\usepackage[round,authoryear]{natbib}
\usepackage{enumitem}
\usepackage{caption}
\usepackage{xurl}
\usepackage[hidelinks]{hyperref}
\usepackage{microtype}
\usepackage{etoolbox}
\usepackage{float}
\usepackage{textcomp}
\usepackage{xcolor}

\graphicspath{{./}}
\modulolinenumbers[1]
\AtBeginEnvironment{tabular}{\singlespacing}
\AtBeginEnvironment{tabularx}{\singlespacing}
\newcommand{\chg}[1]{#1}
\newcommand{\revised}[1]{#1}
\newcommand{\keywords}[1]{\par\noindent\textbf{Keywords:} #1\par}
\newenvironment{abstractblock}{\begin{center}\textbf{Abstract}\end{center}\noindent}{\par}

\begin{document}
\begin{center}
{\LARGE\bfseries Tutor, Not Solver: Designing a Guardrailed AI Assistant for Learning in Higher Education\par}
\vspace{0.75em}
{\large A Design Case of PeteChat\par}
\end{center}
\vspace{1.25em}

{\small
\noindent Belle Li$^{1}$; Lily Tan$^{1}$; Wei Zakharov$^{2,3,*}$; Colby Ben Acton$^{4}$; Qiang Qiu$^{4}$

\medskip

$^{1}$Department of Learning Design and Technology, Purdue University, West Lafayette, Indiana, USA\\
$^{2}$Purdue Libraries and School of Information Studies, Purdue University, West Lafayette, Indiana, USA\\
$^{3}$School of Engineering Education, Purdue University, West Lafayette, Indiana, USA\\
$^{4}$Elmore Family School of Electrical and Computer Engineering, Purdue University, West Lafayette, Indiana, USA

\medskip
\noindent\textbf{Corresponding author:} 
$^{*}$Wei Zakharov, \href{mailto:wzakharov@purdue.edu}{wzakharov@purdue.edu}
\par}
\medskip

\begin{abstractblock}
Generative artificial intelligence (AI) tutors hold significant promise for higher education, yet designing systems that scaffold learning without undermining academic integrity remains an open design challenge. This paper presents PeteChat, a course-aligned AI tutor developed and piloted at a large U.S. research university and documented through the lens of design-based research (DBR). Drawing on formative expert evaluation with teaching assistants and user-experience and developer stakeholders, we report eight transferable design principles for assessment-aware AI tutors, ranging from homework guardrails and debugging scaffolds to self-regulated learning support and instructor-facing customization tools. The system is built on a locally hosted large language model from the Llama-3 family, enhanced with retrieval-augmented generation (RAG) grounded in course-specific materials. Rather than reporting controlled experimental outcomes, this design case foregrounds the situated design reasoning, iterative refinement, and principled decision-making that shaped PeteChat across four development phases. The principles and methodological approach offer actionable guidance for institutions seeking to deploy responsible, integrity-preserving AI tutors at scale.
\end{abstractblock}

\vspace{1em}
\section{Introduction}

Recent advances in generative artificial intelligence have spurred
interest in their applications for higher education. Large Language
Models (LLMs) like OpenAI's GPT-4~\citep{openai2023} are
capable of engaging in human-like dialogue and providing on-demand
information, which presents both opportunities and challenges for
learning environments. Early analyses suggest that such AI tools can
enhance student engagement and provide personalized support, yet they
also raise concerns about accuracy and academic integrity \citep{kasneci2023,yan2023}. Within this context, PeteChat was
conceptualized as a course-aligned LLM-powered learning assistant for
a large U.S. research university. Major design challenges addressed include ensuring
quality of AI-provided information, alignment with course pedagogy,
student data privacy, and upholding academic integrity \citep{cotton2024,slade2013}. By sharing specific details of the
PeteChat implementation --- both successful strategies and pitfalls
encountered --- this design case aims to disseminate actionable design
knowledge that others can build upon \citep{boling2010}.

This paper is structured as a design case \citep{boling2010}, foregrounding
design decisions and rationale over outcome measurements. This format is
appropriate for early-cycle innovations where rich description of the
design process yields transferable knowledge for practitioners facing
similar challenges. Rather than reporting controlled experiments, this
case documents the situated reasoning, iterative refinement, and design
principles that emerged from building and deploying PeteChat in a live
course environment.

\begin{figure}[t]
\centering
\includegraphics[width=0.96\textwidth]{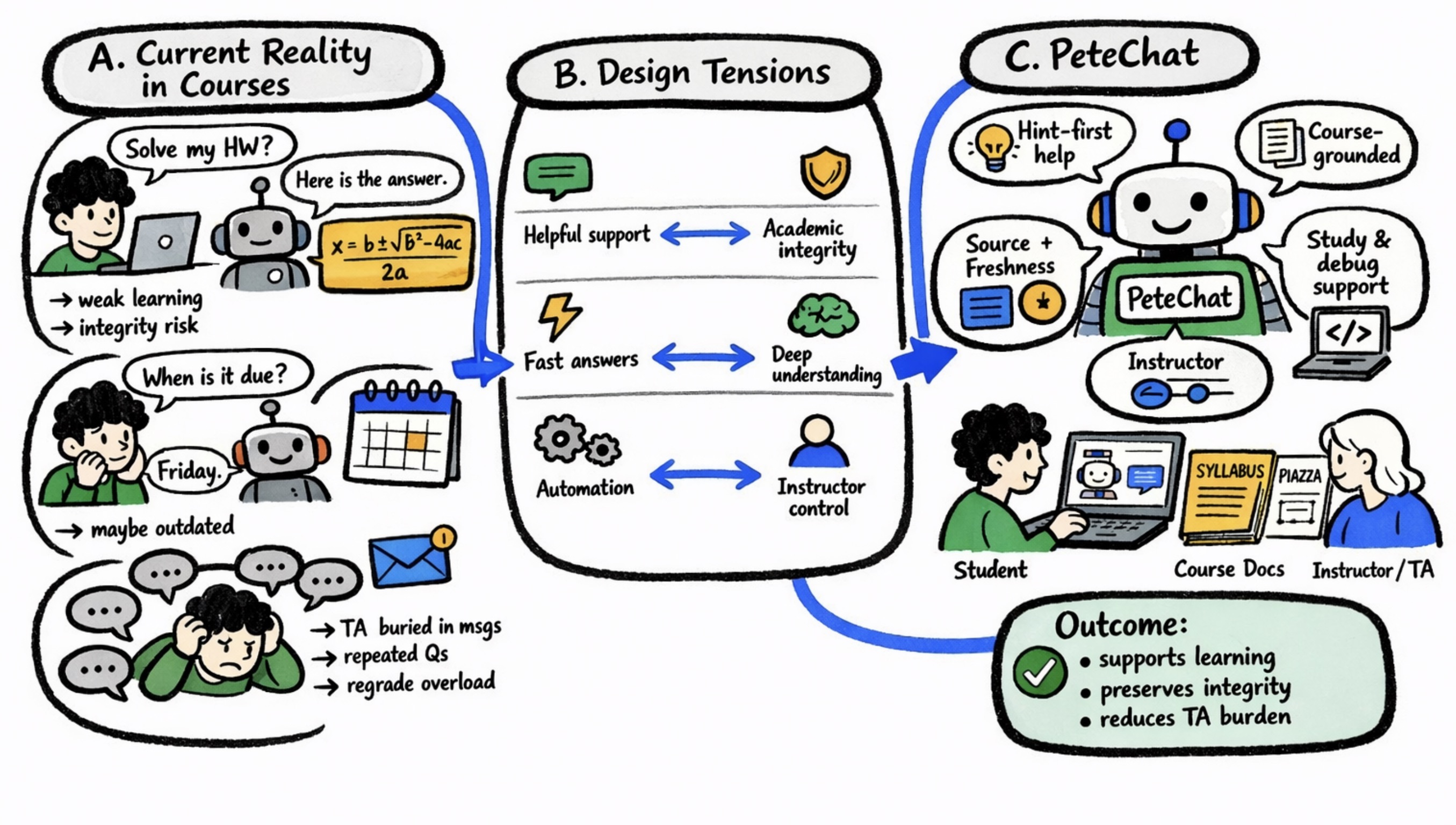}
\caption{\textbf{Conceptual positioning of PeteChat.} The arrows summarize the paper's design rationale: observed course-support problems motivate design tensions, which PeteChat addresses through course grounding, hint-first tutoring, freshness cues, debugging support, and instructor or teaching-assistant controls. The outcome box represents intended design goals rather than measured deployment outcomes.}
\label{fig:petechat-positioning}
\end{figure}

From an HCI perspective, this work contributes to the design of
conversational AI agents in institutional settings --- specifically, how
dialogue constraints, interaction affordances, and information
architecture can be structured to support learning goals within
educational chatbot deployments. The eight design principles reported
here extend prior HCI work on intelligent tutoring system interfaces
\citep[e.g.,][]{kuhail2022} by addressing the underexplored intersection
of academic integrity guardrails, self-regulated learning scaffolds, and
instructor-facing customization in LLM-powered conversational agents.
Onboarding design, response formatting, retrieval provenance display,
and tone configuration are treated here not merely as implementation
\chg{details but as first-class HCI design decisions with practical
implications for student trust and engagement.} Figure~\ref{fig:petechat-positioning}
summarizes the motivating problems, the core design tensions, and the
conceptual positioning of PeteChat as a guardrailed, course-aligned
tutor.

\section{Research Questions}

This study is guided by three questions:

(1) What design principles emerge for AI tutors that scaffold learning
without undermining academic integrity?

(2) How can self-regulated learning (SRL) theory be operationalized in
AI-mediated dialogue?

(3) What institutional constraints shape AI tutor design in practice?

\section{Context and Rationale}

A large U.S. research university, like many institutions, has seen growing student
interest in AI-driven study tools and tutoring systems. The advent of
accessible chatbots (e.g., ChatGPT) created both excitement and
uncertainty among faculty and students regarding their appropriate use
in coursework. On one hand, these tools promise immediate,
context-relevant assistance, such as answering questions about lecture
content or providing feedback on writing, potentially supporting
learners outside of class hours \citep{kasneci2023}. On the other
hand, faculty have expressed concerns about over-reliance on AI or
misuse for completing assignments, which could undermine learning
objectives and violate academic integrity policies \citep{cotton2024}. In response, the PeteChat initiative was launched to proactively
and responsibly integrate AI in a principled way. Rather than banning
AI, the project sought to channel it to support learning: PeteChat was
fine-tuned on curated institutional course materials and aligned with
instructional goals so that it could serve as a ``virtual teaching
assistant'' for students. This fine-tuning of institutional content
aimed to increase the relevance and accuracy of responses for the course
context. By developing our own AI assistant, we could also implement
safeguards and custom behaviors. For example, prompting the model to
refrain from simply giving away homework and assignment answers or to
follow ethical guidelines built into its training data. The rationale
for PeteChat aligns with calls in the literature to harness AI for good
in education while establishing clear frameworks to address its
challenges \citep{kasneci2023}. It also reflects the trend of
universities experimenting with AI copilots for education, recognizing
that outright prohibition is less constructive than guiding appropriate
use.

\subsection{Institutional Context}

The project is supported by an internal university teaching-and-learning innovation program and has been developed
in consultation with the university's teaching-and-learning center, library, and advanced-computing center to
ensure alignment with pedagogy, data governance, and compute resources.

\section{Design-Based Approach}

\begin{figure}[t]
\centering
\includegraphics[width=0.92\textwidth]{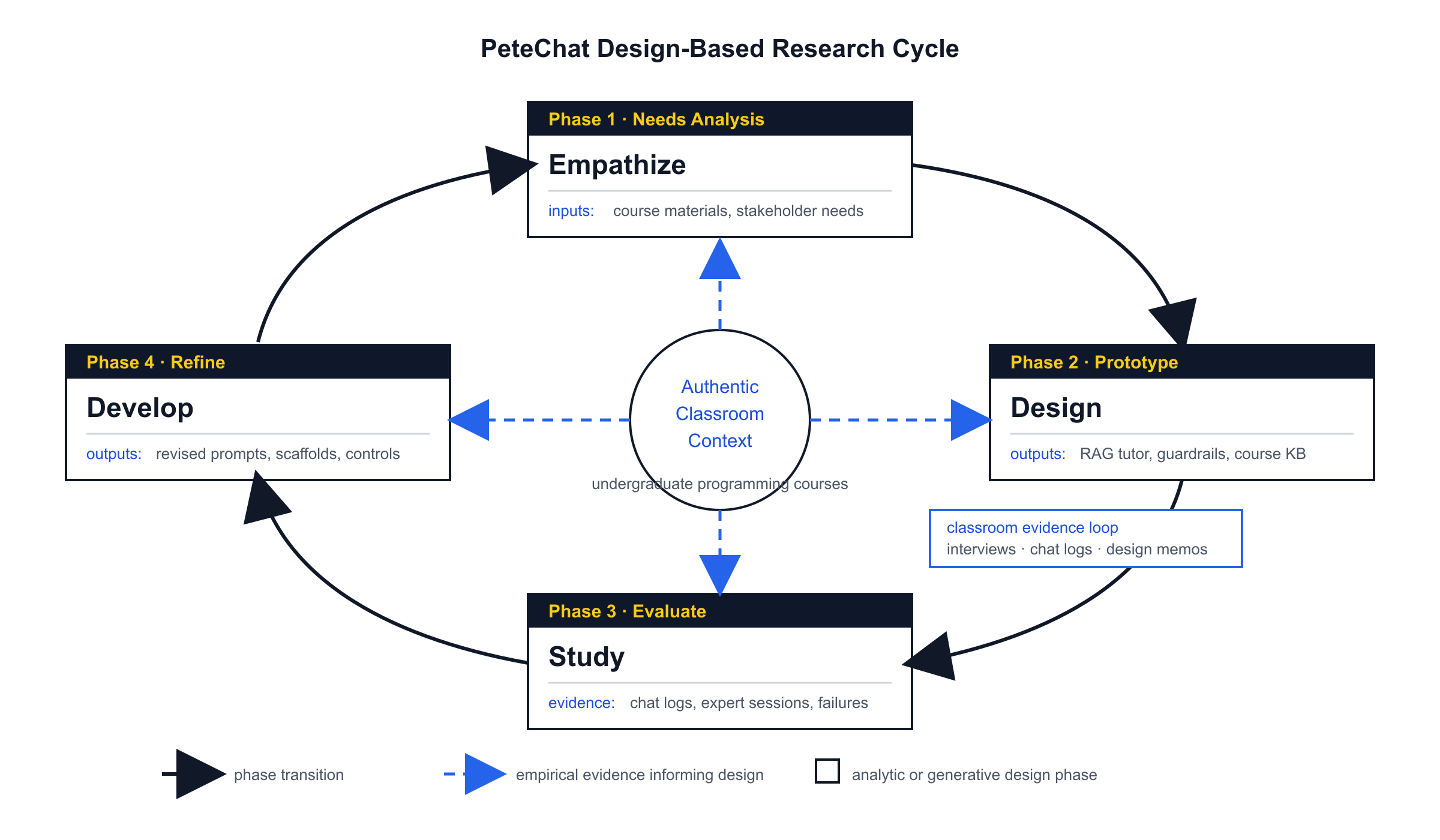}
\caption{\textbf{Design-based research cycle.} Four phases---needs analysis, prototype design, formative evaluation, and refinement---form an outer loop of information flow. Classroom evidence from interviews, chat logs, and design memos informs subsequent redesign.}
\label{fig:dbr-cycle}
\end{figure}

This project was conducted as a design-based research (DBR) study, an
approach well-suited for investigating innovations in real educational
settings through iterative refinement \citep{anderson2012}. In
DBR, researchers and practitioners collaborate to systematically design,
implement, analyze, and re-design an intervention, with the dual goals
of solving a practical problem and generating new knowledge or theory
\citep{dbrc2003}. Unlike controlled laboratory
experiments, DBR emphasizes ecological validity by examining
interventions in authentic contexts and treating them as evolving
designs \citep{barab2004}. We adopted DBR because introducing an
AI assistant into actual courses involves complex interactions
(students, instructors, curriculum, technology) that benefit from
cyclical testing and feedback. Each semester of pilot use constituted a
design cycle in which we collected data on PeteChat's performance and
user experiences, then applied those insights to improve the next
version of the tool. This reflective, iterative process is
characteristic of DBR and helps ensure the innovation is tuned to the
setting's needs \citep{anderson2012}. Figure~\ref{fig:dbr-cycle}
summarizes the four-phase DBR cycle that guided PeteChat's iterative
development and redesign.

Across the project, the work naturally crystallized into four phases,
each building on the lessons of the last:

\subsubsection*{Phase 1 (Summer--Fall 2024)}
\chg{
We began by collecting course-specific data, constructing a baseline
tutor, and fine-tuning open-source LLMs (beginning with Llama-3) on
the university's secure GPU cluster. Internal evaluations (small, rapid, and
exploratory) allowed us to retrain the model quickly as we learned what
worked and what did not.
}

\subsubsection*{Phase 2 (Spring--Fall 2025)}
\chg{
The tutor was deployed through Gradio/Hugging Face and made available to
undergraduates in the focal undergraduate Python course. This marked the first large-scale classroom
test of the system. The Fall 2025 pre-guardrail interaction logs, analyzed
in Section~\ref{sec:research-inputs}, provided our richest window into
how students actually engaged with an AI tutor in a demanding programming
course.
}

\subsubsection*{Phase 3 (Late Fall 2025--Spring 2026)}
\chg{
Using the baseline corpus and expert feedback as design inputs, we began
preference-alignment work to improve the tutor's style and scaffolding
strategies. Because those later alignment and deployment data remain
under analysis, this paper reports them only as design context rather than
as outcome evidence.
}

\subsubsection*{Phase 4 (Spring 2026)}
\chg{
With a redesigned version of PeteChat, the project extended into
additional large undergraduate Python programming courses, marking the
transition from course-specific innovation to broader, campus-level
integration. These deployments are described here as part of the design
trajectory, not as a completed evaluation.
}

\section{Related Work}

Educational chatbot research has grown substantially over the past
decade, providing an increasingly rich design knowledge base. Kuhail et
al. (2022) systematically reviewed 36 educational chatbot studies and
identified that most systems function as teaching agents (delivering
instruction) or peer agents (facilitating collaborative learning), with
the majority relying on predetermined conversational paths and fewer
than a quarter adopting personalized approaches. Their analysis found
that while chatbots generally improved learning outcomes and
satisfaction, persistent challenges included insufficient training data
and neglect of usability heuristics, which are gaps directly addressed
in PeteChat\textquotesingle s design. \citet{labadze2023} extended
this picture through a systematic review of 67 studies, documenting
student benefits (homework assistance, personalized learning, skill
development) and educator benefits (time savings, improved pedagogy),
while highlighting reliability, accuracy, academic integrity, and data
privacy as barriers that careful system design must address.

The emergence of large language models (LLMs) has fundamentally changed
what is possible in educational chatbot design, while simultaneously
intensifying ethical concerns. \citet{kasneci2023} argue that LLMs
hold genuine educational promise, enabling content creation, improved
engagement, and personalized learning, but only if teachers and learners
develop critical competencies to use them responsibly. \citet{cotton2024} sharpen the academic integrity dimension, demonstrating that
ChatGPT substantially lowers the barrier for contract cheating while
making detection far more difficult than traditional plagiarism methods.
Their analysis shows that effective institutional responses require
diversified assessment design, transparent AI policy, and deliberate
student education, which are challenges that directly motivate
PeteChat's guardrailed tutoring approach.

\chg{Prior work also clarifies why institutional AI tutors require more than
technical deployment. A systematic review of AI applications in higher
education found that published work has often been dominated by technical
and STEM-oriented perspectives, with limited attention to pedagogy,
ethics, and educators' roles~\citep{zawacki2019}. For PeteChat, that gap
motivated a design case that documents organizational constraints,
course-level governance, and instructor-facing controls rather than
presenting the chatbot as a purely technical artifact. Work on intelligent
tutoring systems further shows that metacognitive feedback can improve
students' help-seeking behavior when it is embedded in domain learning
tasks~\citep{roll2011}, while research on hypermedia learning highlights
the importance of measuring and scaffolding learners' planning,
monitoring, and reflection processes~\citep{azevedo2010}. \revised{Additional file 1, Table S1}
therefore shifts from review literature to peer-reviewed implementation
studies that report concrete educational AI tutor or assistant systems\revised{~\citep{codeaid2024,codehelp2023,begrading2024,jacobs2024educon}}.
Because the closest published analogues are concentrated in programming
education, the table emphasizes design lessons that transfer to PeteChat:
guardrailed help, classroom deployment, instructor oversight, and
assessment-aware feedback.}

\chg{Self-regulated learning (SRL) theory provides the pedagogical foundation
for PeteChat's dialogue design. In its classic formulation, SRL frames
learning as a cyclical process spanning forethought, performance, and
self-reflection, with learners setting goals, monitoring strategy use,
and evaluating outcomes over time~\citep{zimmerman2002,pintrich2004}.
Learning-science work on metacognitive feedback and hypermedia learning
adds a design caution: students' planning, monitoring, help-seeking, and
reflection should be scaffolded and made observable rather than assumed
from access to an intelligent tool alone~\citep{roll2011,azevedo2010}.
This literature converges on a design imperative motivating PeteChat:
effective AI tutors must constrain themselves to support rather than
replace student cognition, be anchored in course-specific content via
RAG, and scaffold self-regulation explicitly.}

\section{Research Inputs and Pre-Design Analysis}
\label{sec:research-inputs}

Before any prototype was built, we treated PeteChat's design problem as a
pre-design evidence synthesis task. Specifically, we triangulated two
sources of evidence: (1) targeted findings from prior work on
institutional educational chatbots, self-regulated learning (SRL), and
academic integrity, and (2) a directed content analysis of authentic
student--system interactions collected during a pre-guardrail baseline
deployment. This section is therefore not a downstream evaluation of
PeteChat. Instead, it defines the evidence base that shaped the redesign
priorities later formalized in Section~\ref{sec:design-decisions}.

\subsection{Literature-Based Design Inputs}

The targeted literature scan surfaced three recurring design pressures.
First, many prior educational chatbots were only weakly grounded in
course materials or were evaluated in generic, non-institutional
settings~\citep{kuhail2022,labadze2023}. This directly informed our
choice to build PeteChat around course-specific retrieval-augmented
generation (RAG) rather than rely on a generic LLM baseline. Prior
\chg{reviews of educational chatbots further suggested that a retrieval layer
tied to course documents can improve relevance while reducing factual
fabrication by narrowing the assistant's response space to verified
instructional materials~\citep{kuhail2022,labadze2023}.}

\chg{Second, prior work highlighted the need for institutional trust,
privacy, and clear academic-integrity boundaries. Higher education AI
reviews note that institutional adoption raises issues of privacy,
educator role, governance, and ethical risk that cannot be solved by model
accuracy alone~\citep{zawacki2019,yan2023}. Those findings informed our
decision to host PeteChat locally on the university's secure GPU cluster and
to position it explicitly as a learning supplement rather than a solution
engine.}

\chg{Third, SRL literature made clear that effective educational technologies
should support forethought, performance monitoring, help-seeking, and
self-reflection~\citep{zimmerman2002,pintrich2004,roll2011,azevedo2010}.} Taken
together, these studies provided a theoretical blueprint, but they did
not tell us which design failures were most acute in our own classroom
context. For that, we turned to a baseline analysis of authentic
interactions.

\subsection{Pre-Deployment Baseline Analysis}

To complement the literature with institution-specific evidence, we
conducted a directed content analysis of all interactions logged during
a pre-guardrail deployment of PeteChat in the focal undergraduate Python course (Fall 2025). At
this stage, the system operated without the guardrails or SRL scaffolds
described later in Section~\ref{sec:design-decisions}, providing a
naturalistic baseline of student behavior and system response under
largely unmodified conditions. The corpus contains 284 messages from 31
conversations spanning three authentic late-semester task contexts:
exam preparation (47 student / 47 bot messages), homework debugging (50
/ 50), and an open-ended mini-project (45 / 45). Because these messages
were produced under genuine assessment pressure rather than in a
laboratory setting, they offered a practical view of how students and
the baseline system actually behaved.

\chg{No later-phase student outcome data are analyzed in this paper. Usage
logs and preference data from the alignment and cross-course deployment
phases are still being cleaned and reviewed under the project governance
process. We therefore use later phases only to explain design sequencing,
not to claim measured learning gains, adoption rates, or post-redesign
effectiveness.}

We applied a dual-family coding scheme to the full corpus. Category A
(10 codes) captures student SRL-related behaviors grounded in
Zimmerman's three-phase model of forethought, performance, and
self-reflection~\citep{zimmerman2002}. Category B (9 codes) captures
system response alignment with PeteChat's intended design decisions.
Each message received one primary code, with secondary codes permitted
when two behaviors were clearly co-present. The codebook was refined
through pilot coding and revision cycles. To establish inter-rater
reliability, a second independent coder coded a stratified random sample
of 60 messages (21.1\% of the corpus), yielding Cohen's
$\kappa = .82$, indicating strong agreement~\citep{landis1977}.

\begin{figure}[!t]
\centering
\includegraphics[width=0.98\textwidth]{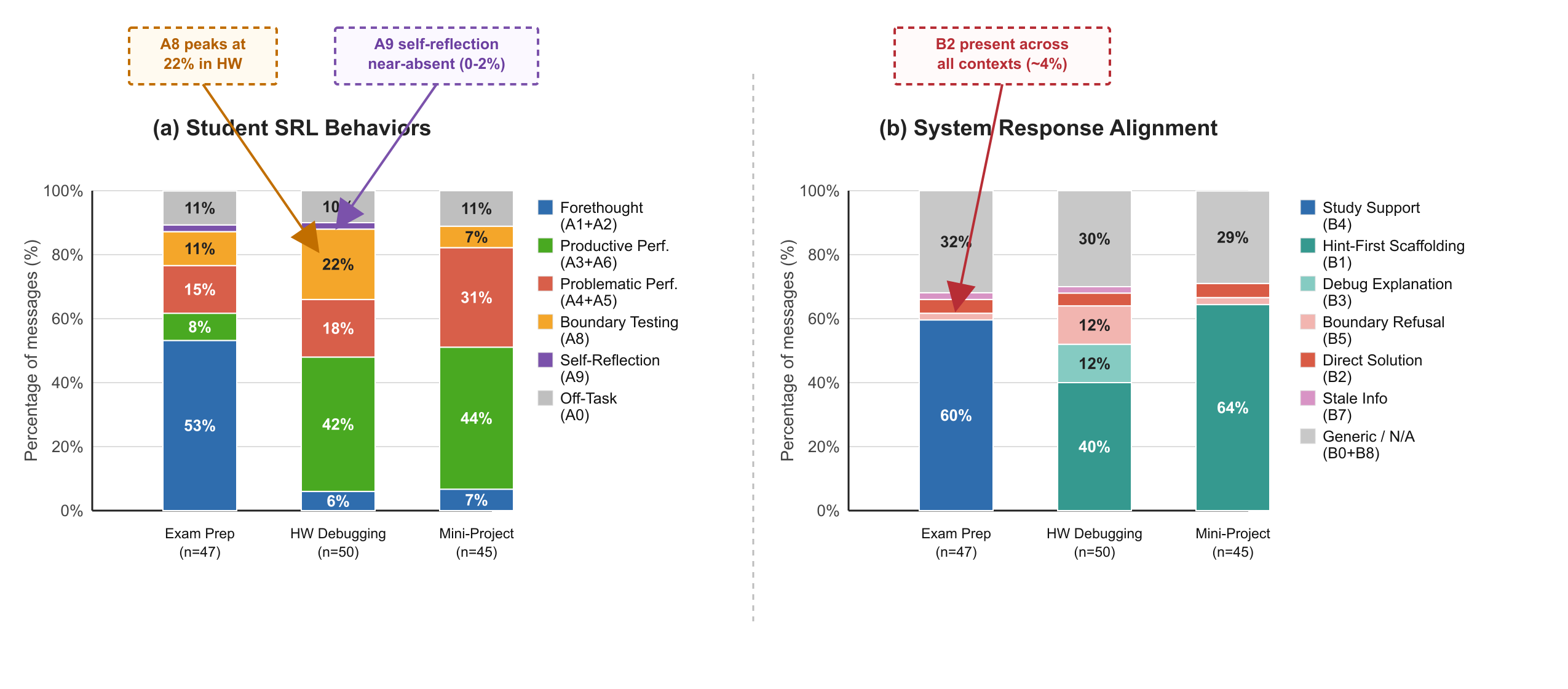}
\caption{\textbf{Baseline interaction patterns across contexts.} Panel (a) shows student self-regulated-learning behaviors across exam preparation, homework debugging, and mini-project contexts. Panel (b) shows system-response alignment across the same contexts. Design-critical patterns include low self-reflection, elevated boundary testing in homework debugging, persistent generic responses, and direct-solution or stale-information failures before guardrails were introduced.}
\label{fig:baseline-cross-context}
\end{figure}

Rather than reproduce the full code distributions in the main text, we
use Figure~\ref{fig:baseline-cross-context} to surface the main
cross-context patterns and defer the complete tables to
\revised{Additional file 1, Section S5 (Tables S2--S4)}. This layout keeps the main text
focused on design-relevant patterns while preserving the full coding
record for inspection.

Three design pressures emerged from the baseline corpus. \emph{First,
boundary and answer-control pressure.} Direct Answer / Solution Seeking
(A5) appeared in 7.0\% of student messages, while Boundary Testing /
System Probing (A8) reached 13.4\% overall and 22.0\% in the homework
context. On the system side, Direct Solution / Over-Solving (B2)
remained present across all three contexts. Together, these patterns
show that students did actively probe for answer-giving behavior and
that the baseline system sometimes complied, establishing an empirical
basis for homework guardrails and stronger assessment-aware routing.

\emph{Second, grounding and freshness pressure.} Although most baseline
responses were not major failures, the corpus still surfaced
Stale or Incorrect Course Information (B7) and a high proportion of
Generic / Non-Instructional output (B0, 29.6\%). These patterns pointed
to two linked deficiencies: the lack of a continuously refreshed
course-grounded retrieval layer for logistics and the absence of
course-specific onboarding or instructor configuration that could shape
response style and relevance.

\emph{Third, SRL visibility pressure.} The baseline corpus showed almost
no observable scaffold uptake or self-reflection. Hint Utilization (A7)
was not observed, and Self-Evaluation / Causal Attribution (A9)
appeared in only two messages (1.4\%), with none in the mini-project
context. In other words, even when students were using the system under
real course pressure, there was little evidence that the interaction was
eliciting the forethought, monitoring, or reflection behaviors that SRL
theory would predict. This made SRL-oriented dialogue scaffolds a design
necessity rather than a merely theoretical aspiration.

\subsection{Convergent Themes and Design Blueprint}

Taken together, the literature review and the baseline analysis
converged on a three-part design blueprint for PeteChat.

\emph{Bounded institutional hosting.} Prior work emphasized privacy,
trust, and academic-integrity concerns, and the baseline corpus showed
that students actively tested system boundaries. These two sources of
evidence jointly motivated a locally hosted, clearly bounded assistant
whose role would be legible as course-aligned tutoring rather than
unrestricted public AI.

\emph{Course-grounded freshness and relevance.} Prior design cases
suggested that RAG improves grounding, while the baseline corpus showed
that stale or generic responses erode confidence. This convergence
motivated a continuously refreshed, course-specific retrieval layer with
source attribution, freshness cues, and instructor-side update paths.

\emph{Explicit SRL scaffolding.} SRL theory argued for goal setting,
monitoring, and reflection, and the baseline corpus showed that such
behaviors were nearly absent without explicit support. This convergence
motivated the goal-setting prompts, hint-first scaffolds, and reflection
micro-checks that later became central to PeteChat's dialogue design.

These patterns did not simply document weaknesses in the baseline
system; they specified what the redesign had to accomplish. The next
section translates these literature-informed and baseline-grounded
pressures into the concrete design decisions that shaped the first
prototype and its subsequent refinements.

\section{First Prototype}

\subsection{System Architecture}

\begin{figure}[t]
\centering
\includegraphics[width=0.98\textwidth]{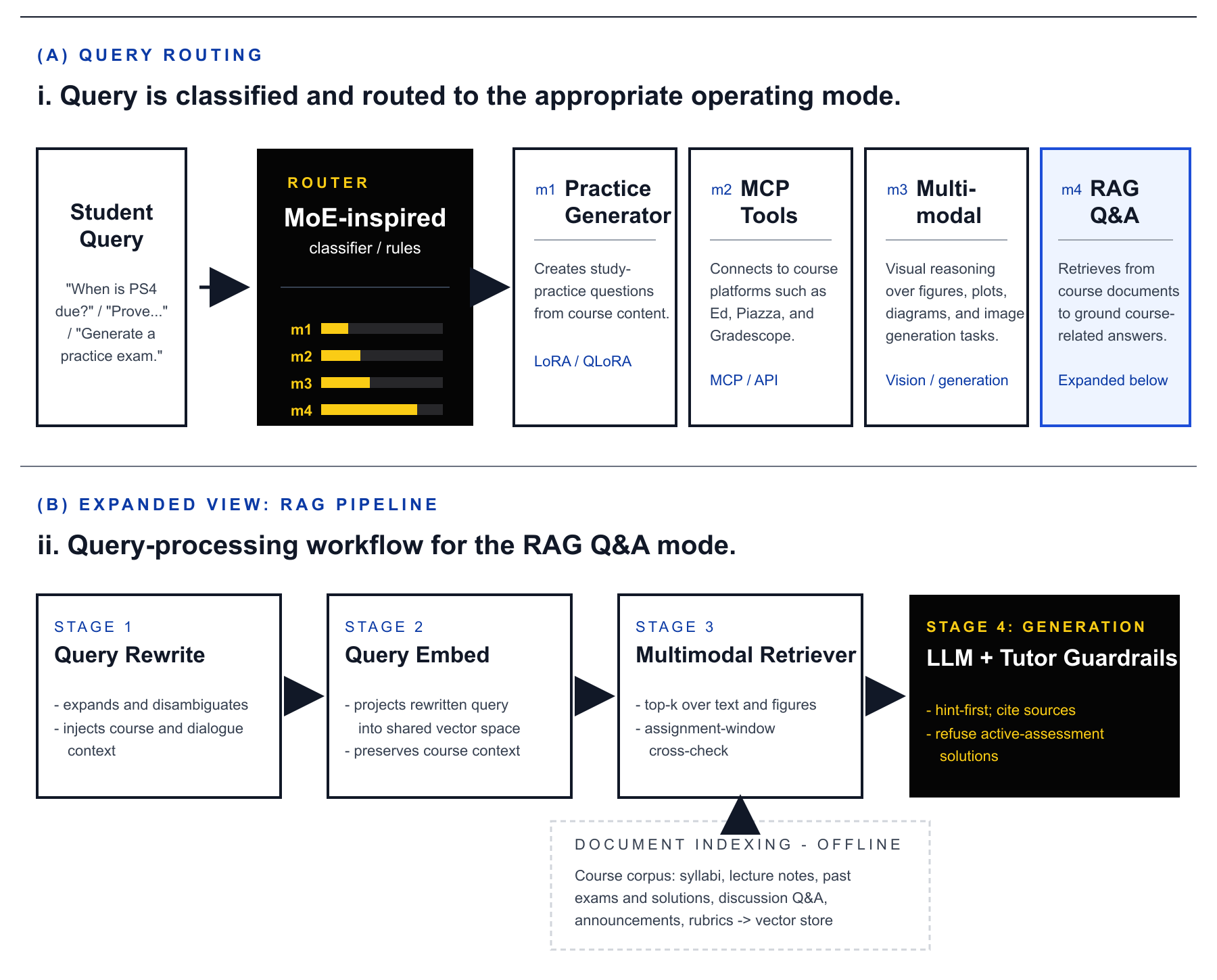}
\caption{\textbf{PeteChat system architecture.} A mixture-of-experts-inspired router dispatches each student query to the most appropriate operating mode: practice-question generation, model-context-protocol tools, multimodal support, or retrieval-augmented-generation question answering. The lower panel expands the retrieval pipeline, including query rewriting, query embedding, multimodal retrieval, and large-language-model generation with tutor guardrails.}
\label{fig:system-architecture}
\end{figure}

PeteChat is built on a locally hosted, open-source large language model
from the Llama-3 family, fine-tuned on institution-specific content using
parameter-efficient fine-tuning (PEFT) techniques, specifically
LoRA/QLoRA, on the university's secure GPU cluster. To tailor the model to our
university context, we assembled a dataset including syllabi, lecture
notes, prior years' exams and solutions, discussion board Q\&As, and
other publicly shared instructional materials from select courses (with
faculty permission). The initial target size (\textasciitilde1,000
instruction-style Q\slash A exemplars in Alpaca format --- a structured
prompt-input-output schema commonly used for supervised fine-tuning of
LLMs) was reached, with incremental augmentation scheduled each semester
thereafter. Fine-tuning aimed to imbue the model with domain knowledge
of course concepts and the preferred problem-solving approaches taught
in those courses. In essence, the AI was trained to become familiar with
``how the focal course teaches a topic,'' enabling it to align its assistance with the
instructors' expectations. Serving was provided via Gradio/Hugging Face
under controlled course access. As a comparative baseline, off-the-shelf
chatbots (e.g., ChatGPT API) were used only to quantify improvements
from local customization; institutional course data were not exported for
third-party fine-tuning. We also established a continuously updated course-scoped knowledge
base (KB) (for example, current semester announcements or
clarifications) that PeteChat could query for the latest information,
addressing the limitation that LLMs have a fixed training cutoff date.
Figure~\ref{fig:system-architecture} provides an overview of the
PeteChat architecture, including the query router and the expanded RAG
pipeline.

\subsection{Dialogue Design}

\begin{figure}[t]
\centering
\includegraphics[width=0.98\textwidth]{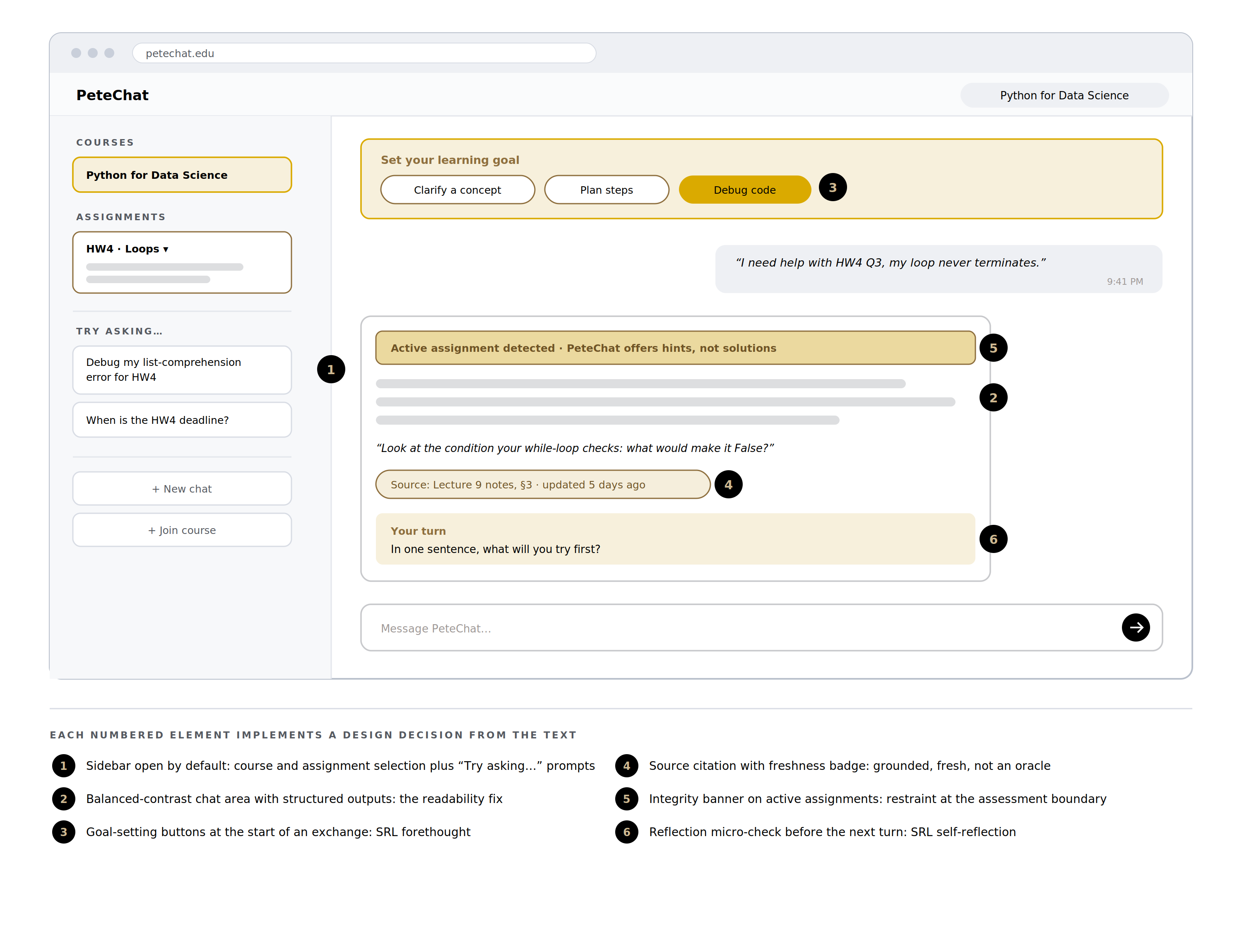}
\caption{\textbf{Annotated first-prototype interface.} Numbered elements connect visible interface choices to the design decisions described in the text: open sidebar prompts, readable structured outputs, goal-setting controls, source-freshness cues, integrity banners, and reflection micro-checks.}
\label{fig:lowfi-mockups}
\end{figure}

A critical aspect of development was crafting the AI's conversational
behavior. We programmed the assistant to follow a tutoring stance: it
tries to guide students to answers with hints and Socratic questioning
rather than directly handing over solutions. This design was influenced
by the need to prevent academic misuse and to encourage learning. For
instance, if asked for homework answers, PeteChat is designed to respond
with clarifying questions or point to relevant resources, nudging the
student to think through the problem. Such decisions were informed by
prior research on intelligent tutoring systems and the known risk of
students using ChatGPT to cheat \citep{cotton2024}. We also
implemented an initial content filter for inappropriate or non-academic
queries, leveraging OpenAI's content moderation tools and custom rules
\chg{during early prototype testing; the course deployment uses a locally}
hosted content filter to maintain student data privacy consistent with
the system's local-hosting rationale. The assistant was tested with
various scenarios to fine-tune its adherence to these guidelines. We
found in early trials that the model sometimes ``hallucinated''
references or overconfidently stated incorrect facts, a known issue with
LLMs \citep{kasneci2023}. To counter this, we adjusted the prompt to
instruct the AI to cite sources from the knowledge base when possible
and to express uncertainty (e.g., saying ``I'm not sure'' or suggesting
where to find an answer) rather than fabricate. While not eliminating
hallucinations entirely, these adjustments reduced blatantly incorrect
answers during our testing phase. Additionally, PeteChat incorporated
reverse prompting and judgment-of-learning (JOL) micro-checks to
encourage metacognition, reinforcing its assessment-aware guardrails.
\chg{Figure~\ref{fig:lowfi-mockups} presents an annotated prototype interface
used to connect concrete interface elements to the design decisions that
emerged from the formative evaluation.}

\section{Evaluation of the First Prototype}

Following the initial deployment of the prototype, we conducted
formative evaluation sessions to identify usability issues, pedagogical
gaps, and opportunities for improvement before the next design
iteration.

Given the formative and early-cycle nature of this evaluation, we
employed a purposive expert sample rather than a large student cohort.
This approach, consistent with DBR practice \citep{anderson2012}, prioritizes actionable insight and rapid iteration over
statistical generalizability.

\subsection{Participants}

We conducted four expert sessions: two TAs for the focal undergraduate Python course (covering
Python, regression, clustering, classification, and neural networks),
one software developer and one UX/UI expert with experience in learning
technologies. Sessions lasted 45--60 minutes each. This sample size is
consistent with expert usability evaluation practice, in which a small
number of evaluators typically surfaces the majority of actionable
usability issues \citep{nielsen1994}. \chg{The developer participant was a
member of the system development team; we therefore treat that session as
stakeholder reflection rather than independent evaluation.}

\subsection{Materials}

Participants interacted with the running prototype (RAG-backed,
course-aware, hint-first) and reviewed sample system outputs.
Semi-structured interview protocols were tailored to each stakeholder
role: TAs focused on student needs and workload reduction, the UX
designer on interface usability and information presentation, and the
developer on technical challenges and alignment concerns.

\subsection{Procedure}

We conducted semi-structured interviews tailored to each stakeholder
role. TA interviews began with warm-up questions about their
responsibilities, and the types of challenges students struggle with
most. Participants then reviewed sample system outputs, rating accuracy
(0--100\%) and identifying helpful versus confusing or misaligned
elements. Follow-up questions explored student needs the prototype did
or did not meet, repetitive TA tasks that PeteChat could realistically
handle, and features that would most reduce workload.

The UX designer interview focused on first impressions of the interface,
expectations versus actual behavior, missing affordances, clarity and
readability of outputs, and ideal redesign priorities. The developer
interview covered the origins of the chatbot concept, early design
goals, decision-making processes during development, technical
challenges encountered, and reflections on which design choices worked
well. All sessions concluded with open-ended prompts inviting any
additional insights. Audio was recorded; transcripts were anonymized and
stored on institutional systems. \chg{The first author conducted an initial thematic
summary of each session, then reviewed the summaries with another member
of the research team to identify convergent design implications. Because
the expert-session sample was small, role-specific, and used primarily
for formative design guidance, we did not compute inter-rater reliability
for these qualitative themes; the reported reliability statistic applies
only to the separately coded baseline interaction corpus in
Section~\ref{sec:research-inputs}. Full interview protocol can be found
\revised{in Additional file 1, Section S2.}}

\section{Findings from Evaluation}

\chg{Analysis of the four expert sessions yielded formative design insights
organized around three stakeholder groups (students, TAs/instructors, and
the development team) with cross-cutting implications for user
experience. These sessions are not presented as a mixed-methods outcome
evaluation; rather, they provide expert and stakeholder input that was
triangulated with the baseline interaction corpus to guide redesign.}

\subsection{Student-Oriented Findings}

\chg{TA participants reported that heavy reliance on generic AI tools like
ChatGPT was inflating homework scores (often above 95\%) while exam
performance dropped, suggesting superficial engagement with course
material.} TAs observed that students frequently paste TA advice directly
into ChatGPT rather than processing it themselves. Additionally,
students increasingly skip critical instructions, such as file-naming
conventions for autograders, because, as one TA put it, they ``are not
reading anymore.''

Participants identified high-value use cases where AI assistance would
be most beneficial: debugging support framed as error-message
explanations rather than code fixes, study planners and mock exams
calibrated to available time, and clarification of assignment intent
when specifications seemed ambiguous.

\subsection{\texorpdfstring{\chg{TA and Instructor-Policy Findings}}{TA and Instructor-Policy Findings}}

TAs identified regrade requests submitted without justification as a
major time sink. Students often do not understand why points were
deducted and submit appeals reflexively. TAs envisioned a system that
could auto-explain point deductions by referencing posted rubrics before
allowing students to escalate to human review.

\chg{TA participants, drawing on instructor-set policies and expectations,
emphasized the need for guardrails: the system should avoid providing
direct homework answers and instead respond with scaffolded, step-by-step
hints accompanied by academic-integrity reminders. For logistics
questions (e.g., deadlines, policies), responses should include ``confirm
with syllabus/Piazza'' disclaimers to protect instructors from liability
when information becomes stale. Participants also recommended per-course
configuration options and alignment audits to ensure the assistant mirrors
the instructor's pedagogical approach.}

\subsection{Developer and UX Findings}

The developer noted that query rewriting and multimodal RAG were already
strengths of the prototype but stressed the importance of minimizing
hallucinations and unfaithfulness to source documents. A consistent
"tutor tone" aligned with course documents was identified as a priority.

The UX designer highlighted critical onboarding gaps: new users did not
know what to ask, the sidebar was closed by default, and the chat
area\textquotesingle s high-contrast styling (white textbox on dark
background) reduced readability. Recommended fixes included an open
sidebar with course-organized ``Try asking\ldots'' example prompts,
lower-contrast chat backgrounds, and visual outputs (charts/tables) for
multi-step answers. An optional share/confirm action was also suggested
so students could verify answers with peers or TAs.

These findings converged on a positioning strategy: PeteChat should be a
``guardrailed, course-aligned tutor'' that reduces TA load and deepens
learning rather than a ``code-dumping bot'' that inflates grades. The
platform\textquotesingle s current strengths, multimodal RAG and query
rewriting, should be preserved while addressing the identified gaps.
These insights directly informed the design decisions for the next
iteration.

\section{Design Decisions Informed by First-Cycle Evaluation}
\label{sec:design-decisions}

Based on the formative evaluation findings, we refined
PeteChat\textquotesingle s design through the following decisions. Each
decision addresses specific gaps or opportunities identified during the
expert sessions.

\textbf{Design Decision 1: Add homework guardrails that refuse direct
solutions and offer scaffolded hints with integrity warnings}

\paragraph{Evidence addressed.} The need for homework guardrails was visible both in the baseline corpus and in the expert sessions. In the baseline analysis, direct answer seeking (A5) and boundary testing (A8) appeared across contexts, while the system still produced occasional direct solutions (B2). TAs then confirmed that heavy GPT reliance is inflating homework scores (north of 95\%) while exam performance drops --- a pattern consistent with broader findings on AI-enabled academic dishonesty \citep{cotton2024}. Together, these findings established that PeteChat needed to refuse direct homework answers while still offering pedagogically useful support.

\paragraph{Design response.} When PeteChat detects a query resembling an
active assignment item, it activates an assessment-aware mode: (a)
refuse to output complete solutions, (b) respond with step-by-step hints
aligned with the instructor\textquotesingle s method, and (c) surface an
academic-integrity banner reminding students of course and university
policies. If the student persists, the assistant offers to explain
underlying concepts or redirect to office hours rather than provide
answers. See Figure 5.

\paragraph{Implementation.} Instructors tag assignments in the
content-upload interface, enabling the system to recognize active items.
A retrieval layer cross-references submission windows with incoming
queries.

\begin{figure}[H]
\centering
\includegraphics[width=0.98\linewidth]{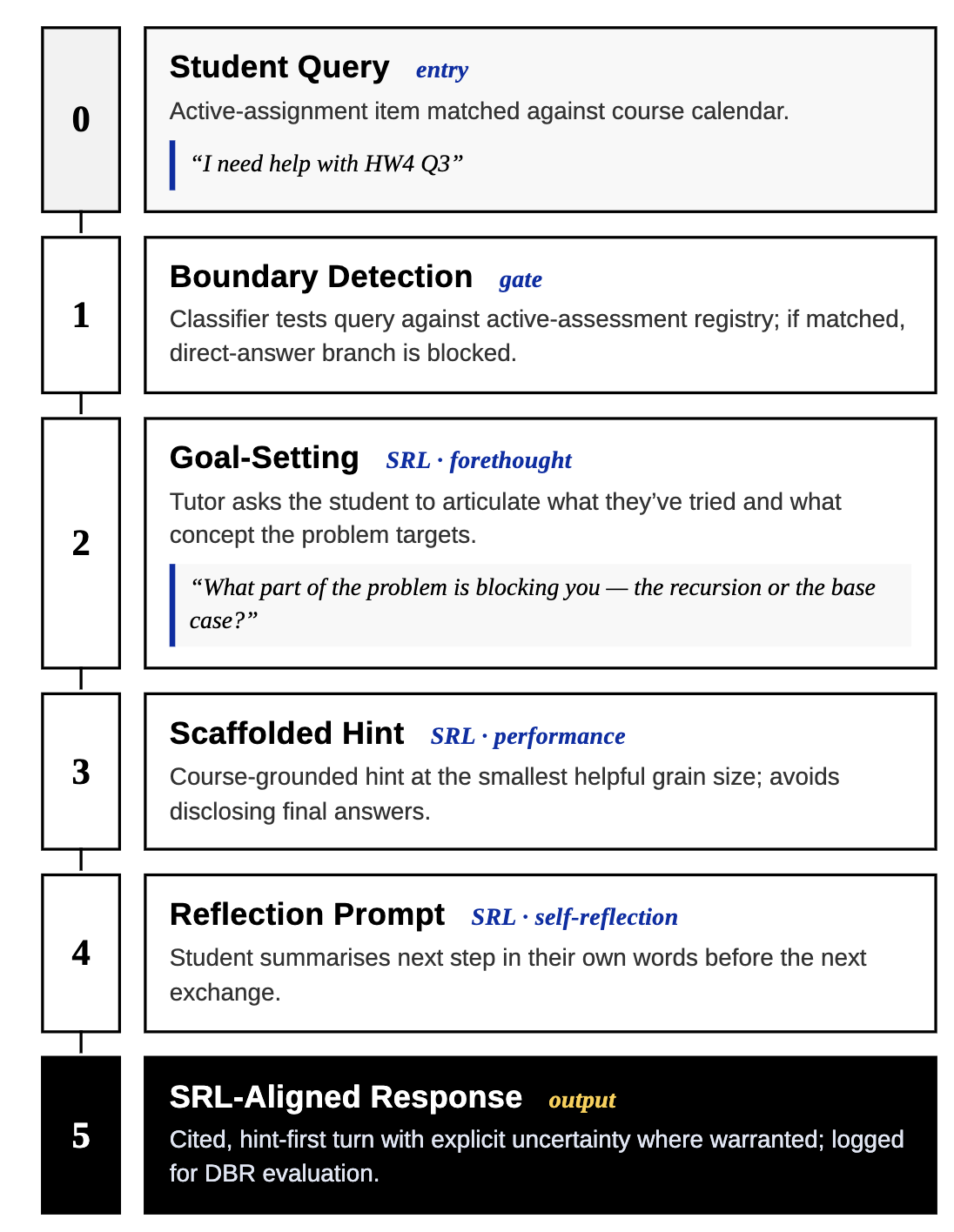}
\caption{\textbf{Assessment-aware interaction flow.} When a student query matches an active assignment, PeteChat routes it through five stages---boundary detection, goal setting, scaffolded hinting, reflection, and a self-regulated-learning-aligned response---instead of providing a direct answer.}
\label{fig:conceptual-interaction-model}
\end{figure}

\textbf{Design Decision 2: Provide logistics answers with freshness
disclaimers and ``check syllabus/Piazza'' prompts}

\paragraph{Evidence addressed.} \chg{The baseline corpus surfaced stale or incorrect course information (B7), and TA participants independently noted that logistics answers should ship with a "check syllabus/Piazza" disclaimer to avoid instructor blame when information becomes stale. Taken together, these findings suggested that logistics support was useful but needed both stronger retrieval freshness and explicit verification cues.}

\paragraph{Design response.} When responding to logistics queries, PeteChat
(a) retrieves the relevant passage from the syllabus or Piazza
announcements, (b) displays a timestamp indicating when the source was
last updated, and (c) appends a disclaimer: "For the most current
information, please verify on Brightspace or Piazza." This positions the
assistant as a first-pass convenience rather than an authoritative
oracle. Figure 6 illustrates how student prompts are routed through
assessment checks, coaching pivots, hint-first responses, reverse
prompting, and judgment-of-learning (JOL) micro-checks to support
scaffolded learning rather than direct answer provision.

\paragraph{Implementation.} Course staff can push updated documents through
a lightweight upload portal that triggers re-indexing within minutes.
The interface badges sources with a freshness indicator (e.g., updated
within 7 days, 30 days, or older).

\begin{figure}[H]
\centering
\includegraphics[width=0.98\linewidth]{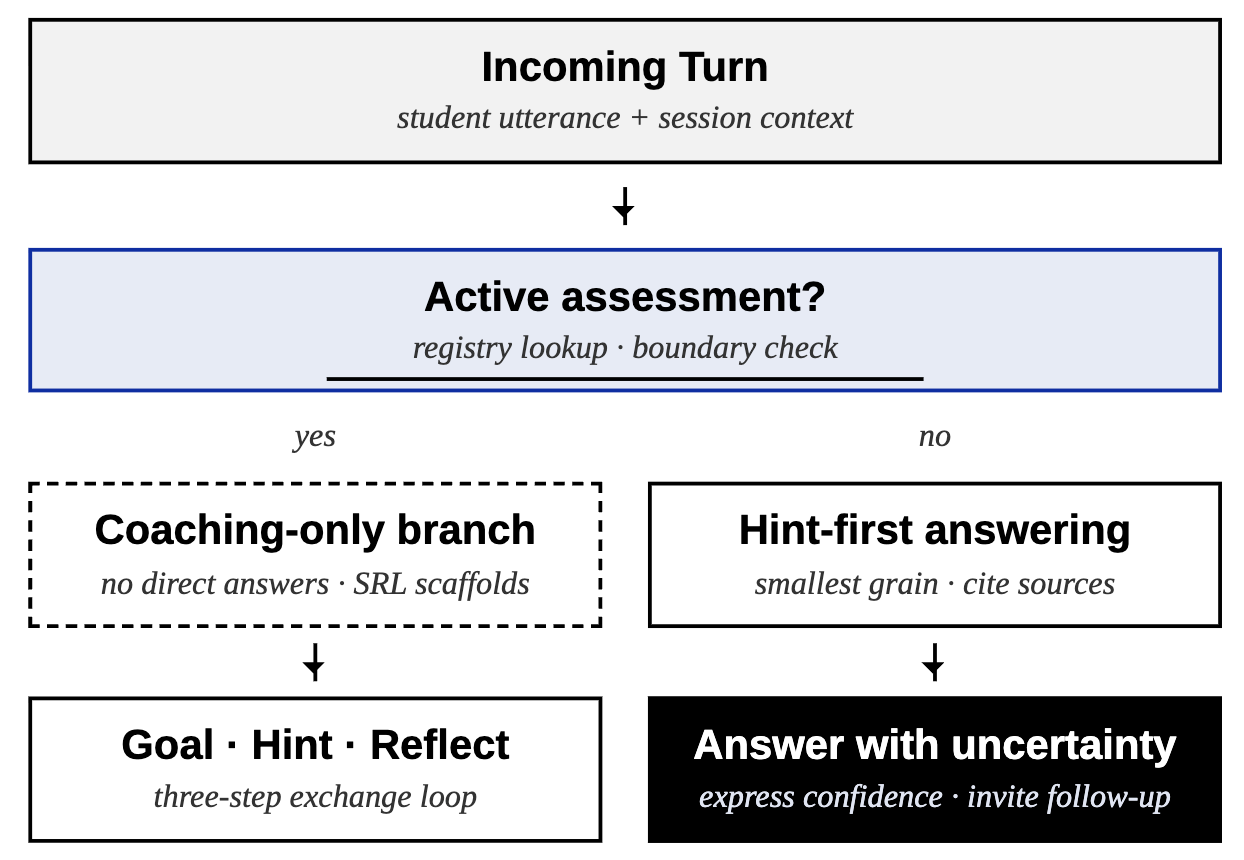}
\caption{\textbf{Turn-level pedagogical decision flow.} Each incoming turn is tested against the active-assessment registry. Matched queries enter a coaching-only branch with goal, hint, and reflection scaffolds; unmatched queries enter a hint-first branch that cites sources and expresses uncertainty.}
\label{fig:pedagogical-dialogue-flow}
\end{figure}

\textbf{Design Decision 3: Frame debugging support as error-message
explanations rather than code fixes}

\paragraph{Interview finding addressed.} Participants identified debugging
support as a high-value use case, but emphasized it should be framed as
error-message explanations that teach students to read and interpret
errors themselves, rather than simply patching code for them. This
reflects a broader finding that chatbots function best as teaching
agents rather than answer providers \citep{kuhail2022}.

\paragraph{Design response.} When students paste error messages or buggy
code, PeteChat (a) explains what the error message means in plain
language, (b) identifies the likely cause without directly fixing the
code, (c) asks the student to articulate what they think is wrong before
offering additional guidance, and (d) provides hints about where to look
in their code. The assistant teaches error-reading skills rather than
providing copy-paste solutions.

\paragraph{Implementation.} A debugging-mode prompt template guides the
interaction. The system detects common Python error types and provides
tailored explanations linked to course materials on debugging
strategies.

\textbf{Design Decision 4: Provide study planners and mock quiz
generators scoped to available time and upcoming exams}

\paragraph{Evidence addressed.} \chg{The baseline corpus showed a striking lack of observable SRL uptake, with hint utilization (A7) absent and self-evaluation (A9) nearly absent, while expert participants identified study plans, mock exams, and time budgets as high-value forms of support that would not undermine academic integrity. This combination made study-planning features an important way to operationalize established SRL constructs such as goal setting, monitoring, and self-evaluation~\citep{zimmerman2002,pintrich2004,azevedo2010}.}

\paragraph{Design response.} PeteChat offers a ``Study Planner'' mode where
students input their available time and target topics. The assistant
generates a time-budgeted study schedule drawing from the syllabus and
past exams. A ``Mock Quiz'' feature generates practice questions from
the course content, provides immediate feedback on answers, and tracks
which topics need more review.

\paragraph{Implementation.} Study planners are auto-generated from syllabus
metadata (exam dates, topic coverage). Mock quizzes draw from a question
bank built from past exams and instructor-provided practice problems,
with randomization to prevent memorization.

\textbf{Design Decision 5: Reduce TA workload through auto-explanations
of point deductions before allowing regrade requests}

\paragraph{Interview finding addressed.} TAs identified regrade requests
without justification as a major workload pain point. Students often do
not understand why points were deducted and submit appeals reflexively.
TAs wanted auto-explanations of point deductions from rubrics.

\paragraph{Design response.} Before students can submit a regrade request,
PeteChat presents a ``Regrade Helper'' flow: (a) retrieve the rubric for
the relevant assignment, (b) explain in plain language which criteria
led to deductions, (c) ask the student to articulate, in their own
words, why they believe the grading was incorrect, and (d) only then
surface the option to escalate to a human TA. This structure filters
unjustified requests while educating students on rubric criteria.

\paragraph{Implementation.} The regrade explanation draws only from
instructor-provided rubrics and solutions, minimizing hallucination
risk. TAs retain full discretion over final grading decisions; the
assistant merely scaffolds the request process.

\textbf{Design Decision 6: Ship a course-specific onboarding pack with a
default-open sidebar and ``Try asking\ldots'' examples}

\paragraph{Interview finding addressed.} UX feedback indicated that
onboarding is unclear. Users did not know what to ask. The sidebar was
closed by default, creating a blank-slate problem --- a gap consistent
with prior findings that educational chatbots often neglect usability
heuristics during design \citep{kuhail2022}. Participants recommended
a default-open sidebar with example questions/topics per course to
reduce friction.

\paragraph{Design response.} The chat interface opens with the sidebar
visible by default, displaying course-organized ``Try asking\ldots''
prompt cards. Examples are tailored to current assignments (e.g.,
``Debug my list-comprehension error for HW2,'' ``Explain the rubric
criteria for Project 1''). Prompt templates cover common categories:
concept clarification, debugging, study planning, and logistics.

\paragraph{Implementation.} Starter prompts are auto-generated from
syllabus metadata (assignment names, due dates, topic tags) and can be
manually overridden by instructors. A usage-analytics layer tracks which
prompts are selected most frequently, feeding back into prompt
refinement.

\textbf{Design Decision 7: Improve visual readability and favor
charts/tables for multi-step answers}

\paragraph{Interview finding addressed.} UX feedback identified
visual/readability issues: the white textbox brightness on dark
background reduced readability. Participants suggested favoring
charts/tables for multi-step answers and adding an optional
share/confirmation feature for answer confidence.

\paragraph{Design response.} Visual contrast on the chat area is reduced to
improve readability. When questions imply multi-step workflows (e.g.,
``How do I preprocess this data?''), PeteChat defaults to graphical or
tabular outputs rather than dense paragraphs. An optional share/verify
action allows students to forward an answer to a peer or TA for
confirmation.

\paragraph{Implementation.} The UI theme is updated with balanced contrast
ratios. Response templates detect multi-step queries and format outputs
as numbered steps, tables, or simple flowcharts. Share links generate a
read-only view of the conversation snippet.

\textbf{Design Decision 8: Provide instructor/TA upload slots, clearer
retrieval provenance, and optional per-course tone configuration}

\paragraph{Interview finding addressed.} The developer expressed concern
about unfaithful retrieval and giving answers too easily. Instructors
wanted a "tutor tone" and alignment with course docs. Participants
requested instructor/TA upload slots for additional documents with
clearer retrieval provenance and optional per-course tone/config.

\paragraph{Design response.} The instructor dashboard includes (a) easy
content upload and update mechanisms for syllabi, rubrics, and
announcements, (b) provenance citations in every response showing which
course documents were retrieved, (c) per-course configuration options
for tone (e.g., more formal vs. conversational) and active-assessment
thresholds, and (d) alignment audits showing how often responses cite
verified sources vs. generate content without grounding.

\paragraph{Implementation.} The upload portal supports drag-and-drop for
common file types. Retrieval provenance is displayed inline in
responses. Tone presets are configurable per course, and alignment
metrics are logged for instructor review.

\section{Design Principles}

The following principles emerged from the synthesis of interview
findings. They guide feature prioritization and resolution of design
trade-offs.

\begin{enumerate}
\def\labelenumi{\arabic{enumi}.}
\item
  \textbf{Tutor, not solver (see Design Decisions 1, 3).} Default to
  hinting and scaffolding; avoid direct answers. Teach how to think
  (debug explanations, step-by-step reasoning).
\item
  \textbf{Align to the course (see Design Decisions 2, 8).} Ground
  responses in instructor-provided materials and show provenance; add
  freshness disclaimers for logistics to protect instructors.
\item
  \textbf{Respect academic integrity (see Design Decision 1).} Include
  guardrails and reminders on homework; encourage students to read and
  follow instructions (e.g., file naming, policies).
\item
  \textbf{Reduce TA overhead (see Design Decision 5).} Automate clarity
  (summaries of instructions), regrade explanations from rubrics, and
  handle repeat questions with consistent, concise answers.
\item
  \textbf{Design for clarity and momentum (see Design Decisions 6, 7).}
  Start with an open sidebar and ``Try asking\ldots'' examples per
  homework/exam; keep answers concise, visual when helpful, and readable
  (balanced contrast).
\item
  \textbf{Flexible control for staff (see Design Decision 8).} Let
  instructors/TAs upload/update content, set tone, and review alignment
  metrics; make it easy to tune without code.
\item
  \textbf{Trust through transparency (see Design Decisions 2, 7).} Show
  sources, note uncertainty, and provide optional share/confirm actions
  so students can verify with peers/TAs.
\item
  \textbf{Time-aware support (see Design Decision 4).} Offer study plans
  and mock quizzes scoped to available time and upcoming exams to keep
  students on track.
\end{enumerate}

\revised{Additional file 1, Section S3 (Figure S1)} shows how several of these principles were instantiated in the second prototype through concrete interface states for both students and instructors, while \revised{Section S4 (Figure S2) of the same file} collects additional screens for debugging, logistics support, study planning, JOL checks, and TA-facing sharing or feedback flows.

\section{Limitations and Future Directions}

\chg{Several limitations should be acknowledged. First, although the paper now combines design documentation with a directed content analysis of a 284-message pre-guardrail baseline corpus and four expert sessions ($n=4$), the overall evidence base remains formative rather than outcome validating; large-scale student learning data from later deployments will still be needed to substantiate learning claims.
Second, both the baseline corpus and the prototype evaluations are situated in a single large research university and primarily in one Python programming course; the baseline analysis in particular reflects one semester, 31 conversations, and one institutional context, so transferability to other disciplines, institutions, or student populations remains to be tested.}
Third, as with all LLM-based systems, PeteChat carries an inherent
hallucination risk --- even with RAG grounding, the model may generate
plausible but incorrect information, particularly for queries outside
the retrieval corpus; provenance citations and uncertainty disclaimers
mitigate but cannot eliminate this risk. Fourth, because the design team
and evaluators overlap substantially, there is potential for self-report
bias; future work should incorporate independent third-party
evaluation. Fifth, the SRL-aligned features (goal-setting prompts,
reflection scaffolds) are theoretically motivated but have not yet been
empirically validated against learning outcome measures. \chg{Sixth, the
alignment and cross-course deployment work in Phases 3 and 4 remains
ongoing; claims about these phases are contextual rather than evaluative
and will require dedicated reporting as results become available.}

\section{Contributions}

Revisiting the three research questions posed at the outset: (RQ1) Eight
transferable design principles emerged for assessment-aware AI tutors
that scaffold learning without undermining academic integrity. (RQ2)
Self-regulated learning theory was operationalized through dialogue
constraints, reflection prompts, and metacognitive scaffolding embedded
across four design decisions. (RQ3) Design-based research proved a
viable and generative methodology for iterative AI tool development in
live institutional settings --- each evaluation cycle directly shaped
subsequent design decisions.

This design case contributes to the growing literature on generative AI in higher education by offering theoretical, design-oriented, and methodological insights into how large language models can be responsibly integrated as course-aligned learning assistants. Rather than evaluating a generic AI tool, this study documents the iterative design of an institutional, assessment-aware AI tutor developed and refined in authentic classroom contexts using a design-based research approach. The contributions are threefold, and they are strengthened by an explicit pre-design baseline analysis that anchors the later design choices in authentic classroom failure patterns rather than only in abstract principles.

First, this study contributes a set of transferable design principles
for assessment-aware, integrity-preserving AI tutors in higher
education. While prior work has documented the affordances and risks of
generative AI for learning, few studies articulate how AI systems can be
deliberately designed to support learning processes without undermining
academic integrity. This design case advances a principled model of an
AI tutor that operates under explicit assessment boundaries. Key design
principles include a hint-first tutoring stance, automated detection of
assessment-relevant prompts, refusal of direct solutions during active
assignments, and redirection toward conceptual explanation, strategy
rehearsal, or human support when appropriate. These principles provide
actionable guidance for institutions seeking alternatives to blanket AI
bans and extend existing discussions of ``responsible AI use'' into
concrete, system-level design patterns.

\chg{The educational contribution is therefore not a new model architecture
alone. It lies in translating familiar institutional concerns --- homework
misuse, stale logistics answers, unproductive help-seeking, regrade
overload, and uneven instructor control --- into course-level design
mechanisms that instructors and TAs can actually govern. PeteChat's value
is organizational as well as pedagogical: the design distributes
responsibility across student-facing scaffolds, instructor-controlled
content updates, provenance displays, and TA-facing review pathways,
making AI adoption less dependent on ad hoc individual prompting.}

Second, the study demonstrates how self-regulated and self-directed
learning theories can be operationalized in AI-mediated dialogue design.
Building on established SRL frameworks \citep[e.g.,][]{zimmerman2002,pintrich2004} and recent conceptual work on AI-supported learning, this study
shows how theoretical constructs, such as goal-setting, monitoring,
judgment of learning, and reflection, can be instantiated within
conversational AI interactions. Design features such as goal-setting
prompt templates, reverse prompting, JOL micro-checks, and
rubric-anchored feedback translate abstract learning theory into
observable dialogue structures. In doing so, the study moves beyond
viewing AI tutors as answer-delivery systems and instead positions them
as learning coaches that scaffold metacognitive processes. This
contribution bridges a gap between learning theory and the practical
design of generative AI systems.

Third, this design case contributes methodological insight into conducting design-based research on generative AI systems in live institutional settings. Methodologically, the study illustrates not only how DBR can be applied to rapidly evolving AI technologies while maintaining ecological validity and ethical safeguards, but also how a pre-guardrail baseline corpus can be used as a design input that links observed classroom failures to subsequent redesign choices. By documenting early expert
evaluations, stakeholder-informed redesign decisions, and the sequencing
of DBR phases across semesters, the paper offers a concrete example of
how institutions can iteratively prototype, evaluate, and govern AI
tutors under real-world constraints such as privacy, data governance,
and instructor accountability. The design case format foregrounds not
only successful design choices but also tensions, trade-offs, and
boundary decisions, contributing to a growing body of methodological
knowledge on studying AI innovations in vivo rather than in isolated
experimental settings.

Lastly, taken as a whole, this paper offers a design-oriented account of
how institutional constraints, baseline classroom evidence, and
iterative prototyping can be brought into one coherent workflow for
educational AI. We hope this design case provides a practical point of
departure for researchers and practitioners seeking to build
course-aligned, integrity-preserving AI tutors in higher education.

\section*{List of abbreviations}
\begin{description}[style=multiline,leftmargin=3.1cm,labelwidth=2.7cm,itemsep=0.2em,parsep=0pt,topsep=0.2em]
\item[AI] Artificial intelligence
\item[DBR] Design-based research
\item[HCI] Human--computer interaction
\item[JOL] Judgment of learning
\item[KB] Knowledge base
\item[LLM] Large language model
\item[LoRA] Low-rank adaptation
\item[MCP] Model Context Protocol
\item[PEFT] Parameter-efficient fine-tuning
\item[QLoRA] Quantized low-rank adaptation
\item[RAG] Retrieval-augmented generation
\item[SRL] Self-regulated learning
\item[TA] Teaching assistant
\item[UI] User interface
\item[UX] User experience
\end{description}

\section*{Declarations}

\noindent\textbf{Competing interests}\\
The authors declare that they have no competing interests.

\medskip
\noindent\textbf{Funding}\\
This project was supported by Purdue University's Innovation Hub Funding Program: Teaching \& Learning in an AI-Rich Environment. The funder had no role in the preparation of this manuscript beyond supporting the project activities described.

\medskip
\noindent\textbf{Authors' contributions}\\
BL, LT, WZ, CBA, and QQ contributed to the conception, design, development, evaluation, interpretation, and preparation of the study. All authors read and approved the final manuscript.

\medskip
\noindent\textbf{Acknowledgements}\\
The authors acknowledge the Center for Instructional Excellence, Purdue Libraries, and the Rosen Center for Advanced Computing for consultation related to pedagogy, data governance, and computing resources.

\medskip
\noindent\textbf{Ethics approval and consent to participate}\\
The study was reviewed and approved by the Purdue University Institutional Review Board (STUDY2026-00000125). All participants provided informed consent prior to the sessions.

\medskip
\noindent\textbf{Availability of data and materials}\\
The datasets generated and/or analyzed during the current study are not publicly available because they contain potentially identifiable student-interaction and interview data and are governed by institutional approvals. De-identified data may be available from the authors on reasonable request and subject to applicable institutional approval.

\section*{Additional file}
\revised{\noindent\textbf{Additional file 1 (included after the references in this preprint).} Supplementary tables, interview protocol, and second-prototype interface states. The file contains the educational AI implementation comparison (Table S1), the thinking-aloud and interview protocol (Section S2), second-prototype interface states (Figures S1--S2), and complete baseline coding distributions and cross-context summaries (Tables S2--S4).}

\clearpage
\setcounter{section}{0}
\setcounter{figure}{0}
\setcounter{table}{0}
\renewcommand{\thesection}{S\arabic{section}}
\renewcommand{\thefigure}{S\arabic{figure}}
\renewcommand{\thetable}{S\arabic{table}}
\renewcommand{\theHsection}{S\arabic{section}}
\renewcommand{\theHfigure}{S\arabic{figure}}
\renewcommand{\theHtable}{S\arabic{table}}

\begin{center}
{\Large\bfseries Additional file 1\par}
\vspace{0.6em}
{\large Supplementary tables, interview protocol, and second-prototype interface states\par}
\vspace{0.8em}
{\bfseries Tutor, Not Solver: Designing a Guardrailed AI Assistant for Learning in Higher Education\par}
{A Design Case of PeteChat\par}
\end{center}
\noindent This supplementary material accompanies the manuscript. Sections S1--S5 preserve the implementation comparison, interview protocol, prototype interface examples, and complete baseline coding tables.

\section{Educational AI Tutor Implementations}
\label{supp:related-work}
Table~\ref{tab:related-ai-assistants} compares the peer-reviewed implementations discussed in Section 5 of the main manuscript.
\begin{table}[H]
\caption{Peer-reviewed educational AI tutor implementations informing PeteChat}
\label{tab:related-ai-assistants}
\centering
\scriptsize
\color{black}
\setlength{\tabcolsep}{3.5pt}
\renewcommand{\arraystretch}{1.18}
\begin{tabularx}{\textwidth}{>{\raggedright\arraybackslash}p{2.45cm} >{\raggedright\arraybackslash}p{1.95cm} >{\raggedright\arraybackslash}p{2.95cm} >{\raggedright\arraybackslash}X >{\raggedright\arraybackslash}p{2.95cm}}
\toprule
\textbf{System / Study} & \textbf{Setting} & \textbf{Implementation Approach} & \textbf{Evaluation / Evidence} & \textbf{Lesson for PeteChat} \\
\midrule
CodeAid \citep{codeaid2024} & Large introductory C and systems programming course & LLM-powered programming assistant that answers conceptual questions, gives pseudocode explanations, and annotates incorrect code without revealing complete code solutions. & Classroom deployment with about 700 students over 12 weeks; analysis included roughly 8,000 usages, weekly surveys, 22 student interviews, and eight educator interviews. & PeteChat should preserve help-seeking value while withholding direct solutions, and should expose transparency and control mechanisms for both students and educators. \\
CodeHelp \citep{codehelp2023} & First-year computer and data science course & LLM-powered help tool with guardrails, instructor-configurable settings, prompting strategies, and instructor views of student help requests. & Deployed with 52 students for 12 weeks; authors analyzed student interactions, student perceptions, instructor reflections, and classroom-use recommendations. & PeteChat should combine guardrailed student help with instructor-facing configuration, monitoring, and review rather than operating as an ungoverned chatbot. \\
BeGrading \citep{begrading2024} & Programming-assignment grading and feedback & Fine-tuned LLM-based grading system trained on real and synthetic programming-assignment data to produce grades and constructive feedback. & Evaluated against Codestral, GPT-4o, Gemini, TA grades, and human evaluation; the authors report reduced grading burden and prompt, objective feedback as intended benefits. & PeteChat's regrade and rubric-support features should remain grounded in instructor-provided rubrics and should frame automation as TA-supportive feedback rather than autonomous grading authority. \\
Tutor Kai \citep{jacobs2024educon} & Introductory object-oriented and functional programming course & Web-based programming practice environment that uses GPT-4 to generate feedback after students execute code, with prompts designed to avoid giving away the solution. & Evaluated with 51 students over one semester; usage logs, student ratings, and task-level feedback analysis showed that most feedback addressed code errors, while incorrect suggestions and occasional code leakage remained concerns. & PeteChat should treat LLM feedback as bounded scaffolding that requires logging, student-facing quality signals, and explicit safeguards against solution disclosure or hallucinated issues. \\
\bottomrule
\end{tabularx}
\color{black}
\end{table}

\clearpage
\section{Protocol Used (Thinking-Aloud + Interview)}
\label{app:protocol}

\subsection{Session Structure (45--60 min)}

\begin{enumerate}
\def\labelenumi{\arabic{enumi}.}
\item
  Welcome \& consent $\rightarrow$ 2) Warm-up $\rightarrow$ 3) Five think-aloud tasks on the
  live prototype $\rightarrow$ 4) Low-fi walkthrough of future features $\rightarrow$ 5)
  Semi-structured interview $\rightarrow$ 6) Debrief.
\end{enumerate}

\subsection{Moderator Prompts During Tasks}

\begin{itemize}
\item
  ``What are you expecting here?'' / ``What would you do next?'' / ``Say
  what's confusing (if anything).''
\end{itemize}

\subsection{Interview Guide (Post-Task)}

\begin{itemize}
\item
  \emph{Usability \& clarity}: discoverability of actions; navigation
  burden; error prevention/recovery.
\item
  \emph{Pedagogical fit}: adequacy of hints; alignment with course
  methods; independence vs. over-scaffolding.
\item
  \emph{Integrity}: banner clarity; perceived fairness; escalation
  paths.
\item
  \emph{RAG trust}: relevance, redundancy, and how citations affect
  acceptance.
\item
  \emph{Future features}: goal-setting, reflection, dashboard---value
  and priority.
\end{itemize}

\subsection{\texorpdfstring{\chg{Data Handling}}{Data Handling}}
Audio/screen were recorded; notes and
transcripts were anonymized and stored on institutional systems. Analytic memos
and affinity maps were versioned in the project repository; only
aggregate insights appear in dashboards and reports, in line with the
grant's privacy and locality commitments.

\clearpage
\section{Second-Prototype Interface States}
\label{app:second-prototype-screens}

\begin{figure}[H]
\centering
\includegraphics[width=0.96\textwidth,trim=4 4 4 4,clip]{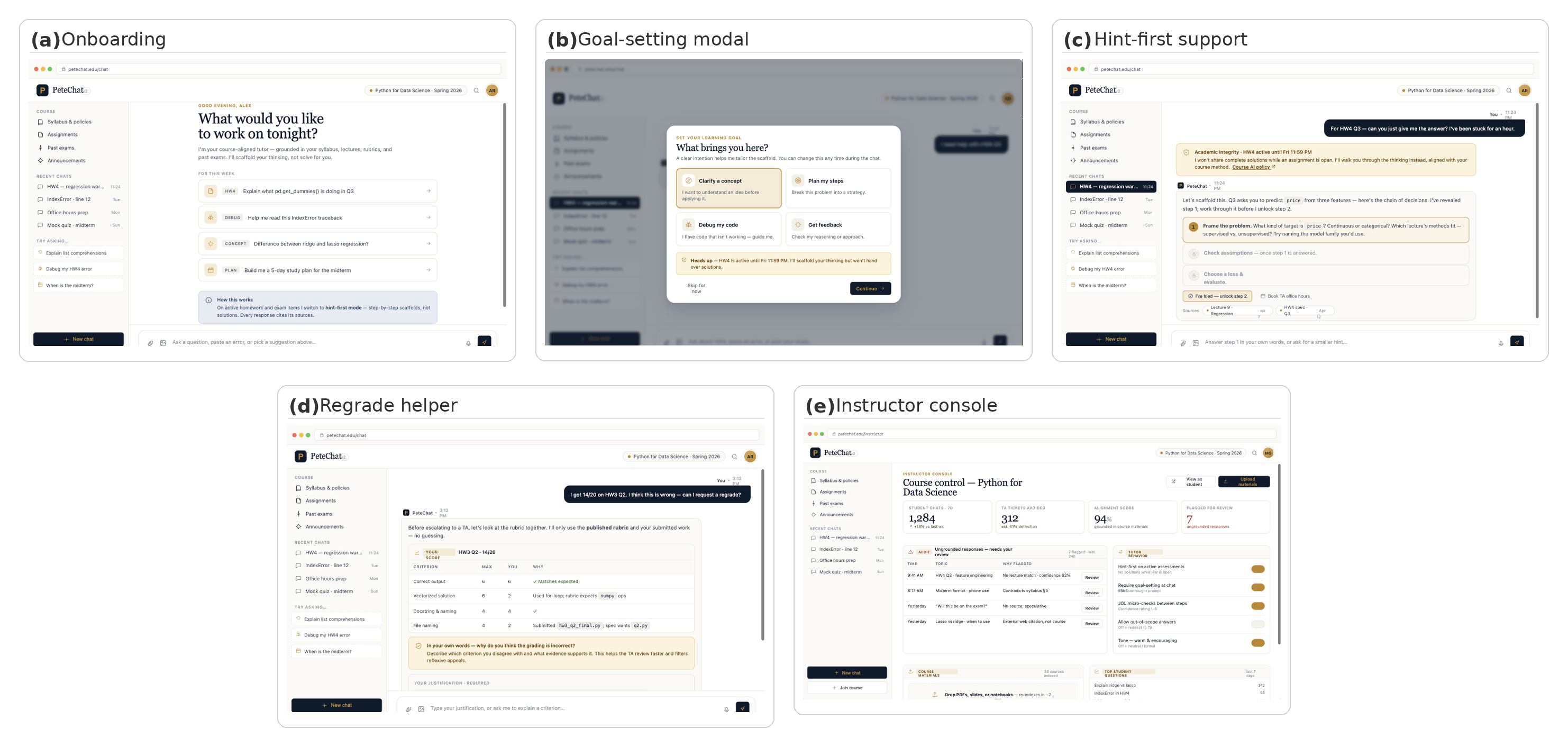}
\caption{\textbf{Second-prototype interface states.} The screens illustrate (a) onboarding prompts, (b) a goal-setting modal, (c) hint-first homework support for active assignments, (d) a regrade helper grounded in published rubrics, and (e) instructor-facing alignment and course-configuration controls. Numeric values are illustrative interface placeholders rather than reported deployment outcomes.}
\label{fig:second-prototype-screens}
\end{figure}

Figure~\ref{fig:second-prototype-screens} presents five representative second-prototype interface states that instantiate the main design principles in concrete UI form: onboarding prompts, a goal-setting modal, hint-first homework support, a rubric-grounded regrade helper, and instructor-facing alignment controls.

\clearpage
\section{Additional Second-Prototype Screens}
\label{app:additional-screens}

\begin{figure}[H]
\centering
\includegraphics[width=0.96\textwidth,trim=4 4 4 4,clip]{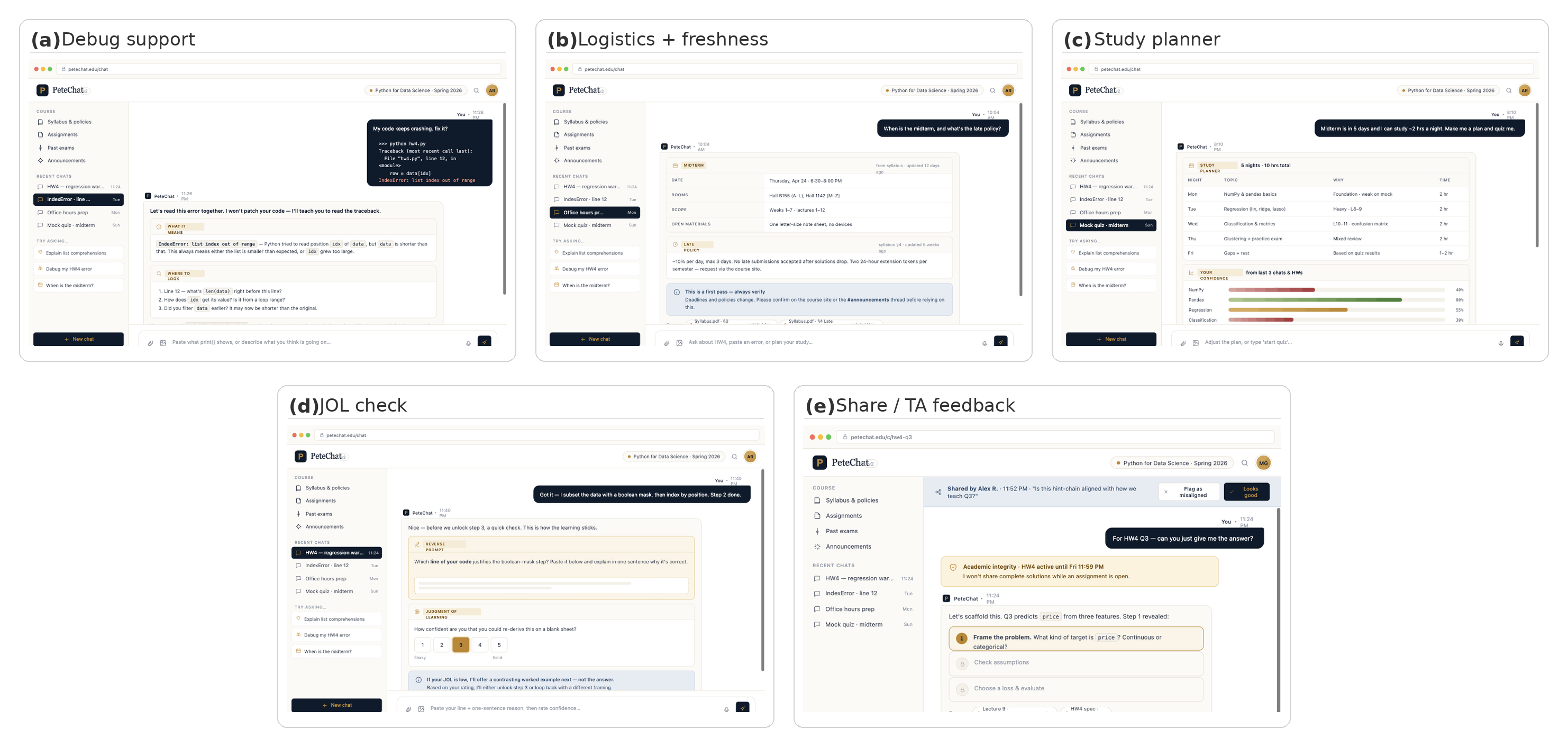}
\caption{\textbf{Additional second-prototype interface states.} The screens illustrate (a) debugging support, (b) logistics answers with freshness disclaimers, (c) a study planner, (d) a judgment-of-learning check, and (e) a sharing or teaching-assistant-feedback view for alignment review. Dates, counts, confidence values, and other numerical details are illustrative placeholders rather than reported outcomes.}
\label{fig:appendix-prototype-screens}
\end{figure}

Figure~\ref{fig:appendix-prototype-screens} collects additional second-prototype screens that complement the interface states in Figure~\ref{fig:second-prototype-screens}. These examples show how the same design principles were extended to debugging support, logistics answers with freshness warnings, study planning, judgment-of-learning checks, and TA-facing sharing or alignment feedback.

\clearpage
\section{Pre-Deployment Baseline Corpus Tables}
\label{app:baseline-tables}

This section reports the complete coding distributions referenced in
Section 6 of the main manuscript. Tables~\ref{tab:student-srl-distribution}--\ref{tab:cross-context-summary}
provide the full Category A and Category B counts and the cross-context
mapping from observed indicators to the design decisions they
motivated.

\begin{table}[H]
\caption{Student self-regulated-learning behaviors by task context}
\label{tab:student-srl-distribution}
\centering
\small
\setlength{\tabcolsep}{5pt}
\renewcommand{\arraystretch}{1.2}
\begin{tabularx}{\textwidth}{>{\raggedright\arraybackslash}p{0.9cm} >{\raggedright\arraybackslash}p{3.0cm} >{\raggedright\arraybackslash}p{2.0cm} >{\raggedright\arraybackslash}p{1.6cm} >{\raggedright\arraybackslash}X}
\toprule
\textbf{Code} & \textbf{Label} & \textbf{SRL Phase} & \textbf{Overall n (\%)} & \textbf{By Context (exam / hw / project)}\\
\midrule
A0 & Off-Task / Non-Learning Use & Off-Task & 15 (10.6\%) & 5 (10.6\%) / 5 (10.0\%) / 5 (11.1\%)\\
A1 & Goal Setting / Task Framing & Forethought & 18 (12.7\%) & 18 (38.3\%) / 0 (0\%) / 0 (0\%)\\
A2 & Strategic Planning & Forethought & 13 (9.2\%) & 7 (14.9\%) / 3 (6.0\%) / 3 (6.7\%)\\
A3 & Adaptive Help-Seeking & Performance & 26 (18.3\%) & 3 (6.4\%) / 14 (28.0\%) / 9 (20.0\%)\\
A4 & Excessive / Broad Help-Seeking & Performance & 20 (14.1\%) & 4 (8.5\%) / 5 (10.0\%) / 11 (24.4\%)\\
A5 & Direct Answer / Solution Seeking & Performance & 10 (7.0\%) & 3 (6.4\%) / 4 (8.0\%) / 3 (6.7\%)\\
A6 & Self-Monitoring / Understanding Check & Performance & 19 (13.4\%) & 1 (2.1\%) / 7 (14.0\%) / 11 (24.4\%)\\
A7 & Hint Utilization / Scaffold Building & Performance & 0 (0.0\%) & 0 / 0 / 0\\
A8 & Boundary Testing / System Probing & Auxiliary & 19 (13.4\%) & 5 (10.6\%) / 11 (22.0\%) / 3 (6.7\%)\\
A9 & Self-Evaluation / Causal Attribution & Self-Reflection & 2 (1.4\%) & 1 (2.1\%) / 1 (2.0\%) / 0 (0\%)\\
\bottomrule
\end{tabularx}
\par\vspace{0.4em}{\footnotesize\emph{Note:} Highlighted rows in the manuscript discussion (A5, A7, A8, A9) indicate design-critical behaviors: direct answer-seeking, absent scaffold uptake, boundary testing, and near-absent self-reflection. $n=142$ student messages across 31 conversations.\par}
\end{table}

\clearpage
\begin{table}[H]
\caption{System-response alignment by task context}
\label{tab:system-alignment-distribution}
\centering
\small
\setlength{\tabcolsep}{5pt}
\renewcommand{\arraystretch}{1.2}
\begin{tabularx}{\textwidth}{>{\raggedright\arraybackslash}p{0.9cm} >{\raggedright\arraybackslash}p{3.3cm} >{\raggedright\arraybackslash}p{2.0cm} >{\raggedright\arraybackslash}p{1.6cm} >{\raggedright\arraybackslash}X}
\toprule
\textbf{Code} & \textbf{Label and Design Decision} & \textbf{Alignment} & \textbf{Overall n (\%)} & \textbf{By Context (exam / hw / project)}\\
\midrule
B0 & Not Applicable / Generic Response & N/A & 42 (29.6\%) & 15 (31.9\%) / 15 (30.0\%) / 12 (26.7\%)\\
B1 & Hint-First Scaffolding (DD1, aligned) & Aligned & 49 (34.5\%) & 0 (0\%) / 20 (40.0\%) / 29 (64.4\%)\\
B2 & Direct Solution / Over-Solving (DD1, misaligned) & Misaligned & 6 (4.2\%) & 2 (4.3\%) / 2 (4.0\%) / 2 (4.4\%)\\
B3 & Debugging-as-Explanation (DD3, aligned) & Aligned & 6 (4.2\%) & 0 (0\%) / 6 (12.0\%) / 0 (0\%)\\
B4 & Study Support / Planning (DD4, aligned) & Aligned & 28 (19.7\%) & 28 (59.6\%) / 0 (0\%) / 0 (0\%)\\
B5 & Boundary Enforcement / Refusal (DD5, aligned) & Aligned & 8 (5.6\%) & 1 (2.1\%) / 6 (12.0\%) / 1 (2.2\%)\\
B7 & Stale / Incorrect Course Info (DD2, misaligned) & Misaligned & 2 (1.4\%) & 1 (2.1\%) / 1 (2.0\%) / 0 (0\%)\\
B8 & No / Empty Response & N/A & 1 (0.7\%) & 0 / 0 / 1 (2.2\%)\\
\bottomrule
\end{tabularx}
\par\vspace{0.4em}{\footnotesize\emph{Note:} Highlighted rows in the manuscript discussion indicate guardrail failures (B2, B7) or generic non-instructional output (B0) motivating design revision. Overall alignment: aligned 91 (64.1\%), misaligned 8 (5.6\%), N/A 43 (30.3\%). $n=142$ bot messages.\par}
\end{table}

\clearpage
\begin{table}[H]
\caption{Cross-context behavioral and alignment summary}
\label{tab:cross-context-summary}
\centering
\small
\setlength{\tabcolsep}{5pt}
\renewcommand{\arraystretch}{1.2}
\begin{tabularx}{\textwidth}{>{\raggedright\arraybackslash}p{3.9cm} >{\centering\arraybackslash}p{1.75cm} >{\centering\arraybackslash}p{1.75cm} >{\centering\arraybackslash}p{1.75cm} >{\raggedright\arraybackslash}X}
\toprule
\textbf{Indicator} & \textbf{Exam Prep ($n=47$)} & \textbf{HW Debug ($n=50$)} & \textbf{Mini-Project ($n=45$)} & \textbf{Design Decision Triggered}\\
\midrule
\multicolumn{5}{l}{\textbf{Student Behavior (Category A): percentage of student messages per context}}\\
A1 Goal Setting & 38\% & 0\% & 0\% & $\rightarrow$ DD4 study planner\\
A3 Adaptive Help-Seeking & 6\% & 28\% & 20\% & $\rightarrow$ DD3 debugging scaffold\\
A4 Broad Help-Seeking & 9\% & 10\% & 24\% & $\rightarrow$ DD3, DD6 onboarding\\
A5 Direct Answer Seeking & 6\% & 8\% & 7\% & $\rightarrow$ DD1 guardrails\\
A6 Self-Monitoring & 2\% & 14\% & 24\% & $\rightarrow$ DD4 SRL scaffolds\\
A8 Boundary Testing & 11\% & 22\% & 7\% & $\rightarrow$ DD5 integrity enforcement\\
A9 Self-Reflection & 2\% & 2\% & 0\% & $\rightarrow$ DD3, DD4 SRL scaffolds\\
\addlinespace[0.35em]
\multicolumn{5}{l}{\textbf{System Response (Category B): percentage of bot messages per context}}\\
B1 Hint-First Scaffolding & 0\% & 40\% & 64\% & $\rightarrow$ DD1 aligned\\
B2 Direct Solution & 4\% & 4\% & 4\% & $\rightarrow$ DD1 guardrails needed\\
B3 Debugging-as-Explanation & 0\% & 12\% & 0\% & $\rightarrow$ DD3 needed\\
B4 Study Support & 60\% & 0\% & 0\% & $\rightarrow$ DD4 aligned\\
B5 Boundary Enforcement & 2\% & 12\% & 2\% & $\rightarrow$ DD5 aligned\\
B7 Stale Information & 2\% & 2\% & 0\% & $\rightarrow$ DD2 needed\\
B0 Generic / Non-instructional & 32\% & 30\% & 27\% & $\rightarrow$ DD6, DD8 needed\\
\bottomrule
\end{tabularx}
\par\vspace{0.4em}{\footnotesize\emph{Note:} This table is intended as a reference companion to Section 10 of the main manuscript. DD1--DD8 refer to the eight design decisions documented in that section.\par}
\end{table}

\clearpage
\end{document}